\documentclass[conference]{IEEEtran}
\IEEEoverridecommandlockouts
\usepackage[dvipsnames,svgnames]{xcolor}

\usepackage{cite,booktabs,array,amsmath,amssymb,amsfonts,tikz,hyperref,wasysym,textcomp,graphicx,algorithmic}  
\usepackage[export]{adjustbox} 
\usepackage[shortlabels]{enumitem}
\usepackage[most]{tcolorbox}

\newcommand{\Sec}[1]{\S{\ref{#1}}}

\usepackage{listings}

\definecolor{codegreen}{rgb}{0,0.6,0}
\definecolor{codegray}{rgb}{0.5,0.5,0.5}
\definecolor{codepurple}{rgb}{0.58,0,0.82}
\definecolor{backcolour}{rgb}{0.95,0.95,0.92}

\lstdefinelanguage{Ini}
{
    basicstyle=\ttfamily\small,
    columns=fullflexible,
    morecomment=[s][\color{red}\bfseries]{[}{]},
    morecomment=[l]{\#},
    morecomment=[l]{;},
    commentstyle=\color{codegray}\ttfamily,
    morekeywords={},
    otherkeywords={=,:},
    keywordstyle={\color{codegreen}\bfseries}
}

\lstdefinestyle{mystyle}{
    backgroundcolor=\color{backcolour},   
    commentstyle=\color{codegreen},
    keywordstyle=\color{magenta},
    numberstyle=\tiny\color{codegray},
    stringstyle=\color{codepurple},
    basicstyle=\ttfamily\footnotesize,
    breakatwhitespace=false,         
    breaklines=true,                 
    captionpos=b,                    
    keepspaces=true,                 
    numbers=left,                    
    numbersep=5pt,                  
    showspaces=false,                
    showstringspaces=false,
    showtabs=false,                  
    tabsize=2
}
\usepackage{multirow}

\usepackage{pifont}

\newcommand{\org}{SPEC-RG}
\newcommand{\cn}{continuum}
\newcommand{\Cn}{Continuum}
\newcommand{\cc}{compute continuum}
\newcommand{\Cc}{Compute continuum}
\newcommand{\CC}{Compute Continuum}
\newcommand{\ra}{reference architecture}

\newcommand{\RA}{Reference Architecture}

\newcommand{\designref}[1]{%
            \begin{tikzpicture}[baseline=(char.base)]
              \node[draw,circle,inner sep=0.5pt, fill=black, text=white] (char){\small #1};
            \end{tikzpicture}%
            }

\makeatletter
\newcommand{\linebreakand}{%
  \end{@IEEEauthorhalign}
  \hfill\mbox{}\par
  \mbox{}\hfill\begin{@IEEEauthorhalign}
}
\makeatother

\begin{document}

\title{The \org{} \RA{} for\\ The \CC{}}

\author{
  \IEEEauthorblockN{Matthijs Jansen}
  \IEEEauthorblockA{\textit{Vrije Universiteit Amsterdam}\\
    Amsterdam, The Netherlands \\
    m.s.jansen@vu.nl}
  \and
  \IEEEauthorblockN{Auday Al-Dulaimy}
  \IEEEauthorblockA{\textit{Mälardalen University}\\
    Västerås, Sweden \\
    auday.aldulaimy@mdu.se}
  \and
  \IEEEauthorblockN{Alessandro V. Papadopoulos}
  \IEEEauthorblockA{\textit{Mälardalen University}\\
    Västerås, Sweden \\
    alessandro.papadopoulos@mdu.se}
  \linebreakand
  \IEEEauthorblockN{Animesh Trivedi}
  \IEEEauthorblockA{\textit{Vrije Universiteit Amsterdam}\\
    Amsterdam, The Netherlands \\
    a.trivedi@vu.nl}
  \and
  \IEEEauthorblockN{Alexandru Iosup}
  \IEEEauthorblockA{\textit{Vrije Universiteit Amsterdam}\\
    Amsterdam, The Netherlands \\
    a.iosup@vu.nl}
}

\IEEEaftertitletext{\vspace{-1\baselineskip}}

\maketitle


\begin{abstract}
    As the next generation of diverse workloads like autonomous driving and augmented/virtual reality evolves, computation is shifting from cloud-based services to the edge, leading to the emergence of a cloud-edge \textit{\cc{}}.
    This \cn{} promises a wide spectrum of deployment opportunities for workloads that can leverage the strengths of cloud (scalable infrastructure, high reliability) and edge (energy efficient, low latencies).
    Despite its promises, the \cn{} has only been studied in \textit{silos} of various computing models, thus lacking strong end-to-end theoretical and engineering foundations for computing and resource management across the \cn{}.
    Consequently, developers resort to ad hoc approaches to reason about performance and resource utilization of workloads in the \cn{}.
    In this work, we conduct a first-of-its-kind systematic study of various computing models, identify salient properties, and make a case to unify them under a \textit{\cc{} \ra{}}.
    This architecture provides an end-to-end analysis framework for developers to reason about resource management, workload distribution, and performance analysis.
    We demonstrate the utility of the \ra{} by analyzing two popular \cn{} workloads, deep learning and industrial IoT.
    We have developed an accompanying deployment and benchmarking framework and first-order analytical model for quantitative reasoning of \cn{} workloads.
    The framework is open-sourced and available at \url{https://github.com/atlarge-research/continuum}.
\end{abstract}

\begin{IEEEkeywords}
  \Cc{}, \ra{}, edge computing, resource management, offloading, benchmark
\end{IEEEkeywords}

\section{Introduction}\label{sec:introduction}
\begin{figure}[t]
    \centering
    \includegraphics[width=\linewidth]{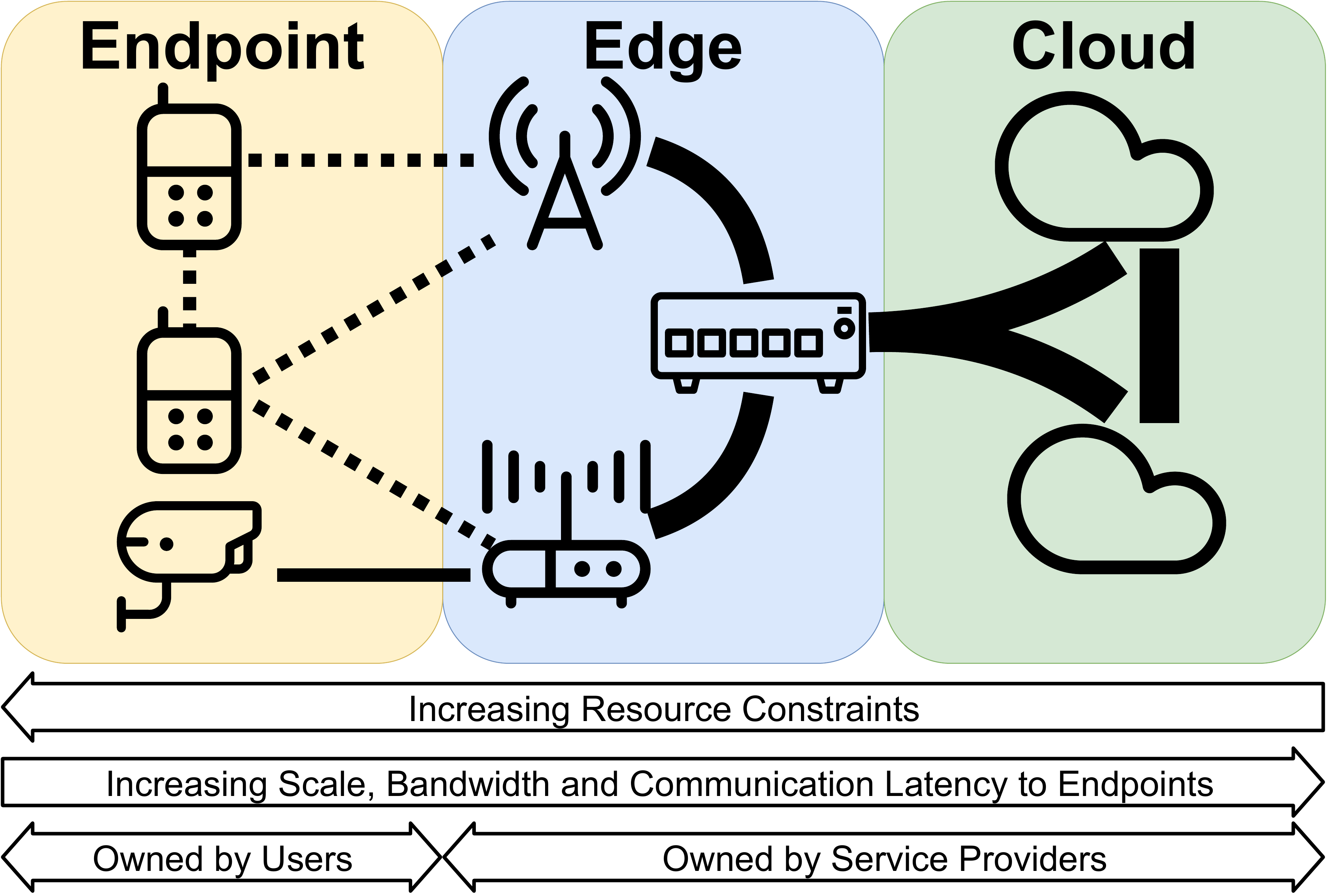}
    \caption{An overview of the \cc{} (key properties shown as arrows at the bottom) with endpoints, edge servers, and cloud infrastructure. }
    \label{fig:sys-model}
\end{figure}

Cloud computing is one of today's most successful computing paradigms~\cite{2021-hotos-sky-computing}. 
Cloud developers can summon a large fleet of servers and infrastructure services (storage, network, resource managers), and deploy complex, scalable workloads on them with a few clicks~\cite{2022-ieee-serverless,2021-cacm-serverless}.
Traditionally, workloads are executed in the cloud or in a limited capacity at the user, located on resource-constrained endpoint devices at the far end of the network such as sensor nodes, smart devices, and IoT devices~\cite{satyanarayanan2019seminal}.
To enable new generations of workloads with strict performance and privacy requirements, the cloud-centric view of computation is shifting outwards towards the \textit{edge}, close to users.
Edge computing offers low latency, energy-efficient data processing by processing data in-the-field, close to where it is generated using decentralized, heterogeneous, and mobile computing devices often with limited resources.
Domains with edge workloads include the Internet of Things (IoT)~\cite{aws2021edge}, self-driving vehicles~\cite{lin2018architectural}, smart farming~\cite{vasisht2017farmbeats}, smart industry~\cite{iiot2018tseng}, mobile gaming~\cite{zhang2019improving}, analytics~\cite{patel2017using}, and machine learning~\cite{khan2019deep}.
With edge computing connecting cloud and user, workloads previously deployed as cloud-only or endpoint-only are now distributed in compute, data, and state and across a \cc{}~\cite{linux2021state} of cloud, edge, and endpoint devices, leveraging the best of both worlds: the high-performance, scalable network-storage infrastructure and high reliability of clouds with low-latency, privacy-preserving computation of edge.
Figure~\ref{fig:sys-model} provides an overview of this \cn.

Though promising, the \cn{} also presents unique challenges to workload developers and infrastructure providers.
Unlike clouds, edge computing lacks standardization of development guidelines and foundational infrastructure services like resource managers, scalable storage, or automatic workflow managers that help with workload deployments~\cite{2020-hotedge-griffin,9355587,2018-hotedge-os-edge,2021-ieee-computing-edge}.
Hence, developers must decide by themselves how to manage resources, split workload, and offload computation with tasks (either in parts or as a whole) that traditionally run in the cloud to edge servers and endpoints~\cite{satyanarayanan2009case}.
However, making such decisions is non-trivial as there is a large design space of choices.
For example, task offloading includes, but is not limited to, offloading from cloud to edge~\cite{xiong2018extend}, offloading from edge to cloud~\cite{kang2017neurosurgeon}, compute management among different edge devices~\cite{mortazavi2017cloudpath}, and compute management between edge and endpoint devices~\cite{satyanarayanan2009case}.

\begin{figure}[t]
    \begin{tcolorbox}[colback=gray!10!white,colframe=black,arc=0mm]
        \textbf{Key insight:} 
        Existing computing models for task offloading are often presented in isolation, while there is significant overlap in concerns addressed by these models.
        By considering cloud, edge, and endpoint computing models as part of a unified, continuous computing model---the \cc{}---developers are no longer limited to the single set of constraints from current isolated computing models.
    \end{tcolorbox}
\end{figure}

\textit{Computing models} like fog, mist, or mobile cloud computing typically address these choices by capturing pertinent workload developments and deployment guidelines for a specific offloading model.
As we will illustrate in \Sec{sec:computing}, such fragmented views lead to \textit{silos} of knowledge and miss an opportunity to develop an \textit{end-to-end} reasoning framework regarding workload splitting, infrastructure services, task offloading, and resource management.
A further consequence of such fragmentation is the lack of any deployment, analysis, and benchmarking framework with strong theoretical and engineering foundations that can help developers assess their decisions' quality (\Sec{sec:relatedwork}).
As a result, \cn{} workload developers make development and deployment decisions ad hoc, thus creating complexity and a general confusion regarding \textit{what is the \cc{}} and \textit{what should a developer or infrastructure provider know about it before developing or supporting compute \cn-ready applications}.

In this paper, we take a step back and systematically study the evolution of the \cc{} and various associated computing models.
Our study of 17 \cc{} models reveals the \textit{primary insight} that even though these computing models are often presented in isolation, there is a significant overlap in terms of concerns addressed, used mechanisms, and application domains addressed by these different computing models.
As a result, we synthesize common characteristics addressed by these models and select five representative models that each cover a distinct part of the \cc{} design space (Table~\ref{tab:computing}): Mist computing~\cite{preden2015benefits}, edge computing~\cite{satyanarayanan2017emergence}, multi-access edge computing~\cite{taleb2017multi}, fog computing~\cite{bonomi2012fog}, and mobile cloud computing~\cite{dinh2013survey}.
Thus, we make a case that developers and infrastructure providers should not consider cloud, edge, and endpoint computing models in isolation but as parts of a unified, \textit{continuous computing model}, captured as the \org{} \cc{} \ra{}\footnote{Established in 2011, the Cloud Group of the SPEC Research Group focuses on general and specific performance issues associated with cloud operation, from traditional to new performance metrics, from workload characterization to modeling, from concepts to tools, from performance measurement processes to benchmarks. The work presented here is part of a larger, long-term activity within this group, focusing on understanding the systems principles of cloud and edge computing. The activity has started in 2019 and has resulted in several publications, which are available online with open-source artifacts. The group agrees with this article's publication under the current co-authorship.}. 
The unified \ra{} identifies the key building blocks of any \cn{} deployment and associated key concerns. 
We map the computing models and two related use cases (deep learning and industrial IoT) to this \ra{} to demonstrate its completeness, comprehensiveness, and usefulness.

The \ra{} also gives us a blueprint to design and implement a workload deployment and benchmarking framework by defining \cc{} components and their responsibility split. 
The framework leverages virtual machines (KVM) and container technologies (Kubernetes) and allows a developer to configure various typical network (bandwidth, latency, jitter, packet drops), storage (capacity, bandwidth), and compute configurations (CPU type, speed, number of cores, memory) found in the \cc{}.
Supporting these configurations with any number of machines allows a developer to quickly explore the performance landscape of various \cc{} deployments in only a few lines of code and collect performance statistics to analyze whether a particular offloading model is suitable for their needs.
The performance reports from the deployment framework are verified using a first-order analytical framework. 
The unique combination of a conceptual model (the Reference Architecture) with practical (deployment and benchmarking framework) and analytical (first-order model) frameworks enables developers to reason about their \cc{} workloads \textit{with} strong theoretical and engineering foundations --- a feat that is not possible today.

Our key contributions in this paper include:
\begin{enumerate}
    \item (\textit{survey}) 
    To the best of our knowledge, we present the most comprehensive survey on computing models with 17 models, synthesize their salient properties, and identify opportunities for unification (\Sec{sec:computing}).
    \item (\textit{conceptual}) 
    We make a case for the \cc{} and propose a design of a unified \ra{}.
    Our \ra{} is the first to consider the \textit{entire edge-cloud \cc{}}.
    We use this new architecture to synthesize two domain-specific architectures for deep learning and industrial IoT (\Sec{sec:refarch}).
    \item (\textit{experimental})
    To aid \cc{} developers, we present an open-source workload deployment and benchmarking framework. 
    We show how developers can explore the large \cn{} design space in a few lines of code to examine various trade-offs (\Sec{sec:results}).
    \item (\textit{analytical}) 
    We enhance and verify the performance analysis capabilities of the framework by formulating an analytical performance model for exploring workload deployment scenarios in the continuum (\Sec{sec:perf_model}).
\end{enumerate}

\begin{table*}[t!]
    \caption{Comparison of Key Characteristics of 5 Computing Models for Task Offloading between Cloud, Edge, and Endpoint. MC: Mist Computing; EC: Edge Computing; MEC: Multi-access Edge Computing; FC: Fog Computing; MCC: Mobile Cloud Computing.}
    \label{tab:computing}
    \centering
    \begin{tabular}{l|lllll}
    \specialrule{.1em}{.0em}{.0em}                                          
    Key Characteristic (KC)              & MC                & EC                    & MEC                    & FC                   & MCC               \\
    \specialrule{.05em}{.0em}{.0em}
    \textbf{KC1:} Compute source         & Endpoint          & Endpoint              & Cloud                  & Cloud                & Endpoint          \\
    \textbf{KC2:} Data source            & Endpoint          & Cloud, endpoint       & Endpoint               & Cloud, endpoint      & Endpoint          \\
    \textbf{KC3:} Offload target         & Endpoint          & Edge                  & Edge                   & Edge                 & Cloud             \\
    \textbf{KC4:} Architecture           & Peer-to-peer      & Distributed           & Distributed            & Hierarchical         & Distributed       \\
    \textbf{KC5:} Offload service        & Compute, storage  & Compute               & Compute                & Compute, storage     & Compute, storage  \\
    \textbf{KC6:} Compute capacity       & Low               & Moderate              & Moderate               & Moderate to high     & High              \\
    \textbf{KC7:} Network latency        & Low               & Low                   & Low                    & Low to moderate      & High              \\
    \textbf{KC8:} Network type           & Wired, wireless   & Wired, wireless       & Wireless               & Wired, wireless      & Wired             \\    
    \textbf{KC9:} Operator               & User              & User, cloud provider  & Network provider       & Cloud provider       & Cloud provider    \\
    \specialrule{.05em}{.0em}{.0em}
    \multirow{2}{*}{Popular use case}    & Peer-to-peer IoT  & Real-time data        & Mobile, network-aware  & Low latency cloud    & Compute- and storage- \\
                                         & processing        & processing            & data processing        & service provisioning & intensive tasks \\
    \specialrule{.1em}{.0em}{.0em} 
\end{tabular}

\end{table*}

\section{A Case for the Unified \Cn{} Model} \label{sec:computing}
Due to the historical and diverse nature of outside-the-cloud computing environments, many detailed but selective computing models have been developed to guide developers and infrastructure providers in deploying and supporting their workload.
Each model has its assumptions on available infrastructure and workload demands, and therefore, finding the right computing model for an application deployment requires a careful analysis of characteristics desired by users and offered by computing models.
In this section, we first synthesize key characteristics of interest to developers and infrastructure providers, then list how various computing models deliver these properties, and finally, make a case for unifying end-to-end concerns.
Past survey works in this field also analyze these models but fall short of providing an end-to-end unification argument~\cite{yousefpour2019all,li2018edgeoriented,mouradian2018comprehensive,taleb2017multiaccess}.

\subsection{Synthesizing Key Characteristics}
We start by synthesizing and identifying 9 unique key characteristics (KC) that a \cc{} developer or infrastructure provider should know about (Table~\ref{tab:computing}).
Based on the computing model and the workload, some or all of these key characteristics can be of interest to a developer or provider.


\begin{enumerate}[label={\textbf{KC\arabic*.}},leftmargin=*]
    \item \textbf{Compute source:} shows where requests for computing are generated.
          The source could be cloud, edge, or endpoint, and is different from the offload target.
    \item \textbf{Data source:} determines where the data is generated or stored before offloading.
    \item \textbf{Offload target:} identifies where computing operations and data are offloaded to in the \cn{}.
    \item \textbf{Architecture:} determines the structure of devices participating in offloading.
          The options are P2P, distributed (a cluster of devices in a single layer of the \cn{}), or hierarchical with multiple sub-layers.
    \item \textbf{Offload service:} captures the type of services being offloaded, e.g., computing requests or data.
    \item \textbf{Compute capacity:} captures the compute capacity of the offload target.
          In a typical setup, endpoint devices have low energy, power-efficient CPUs (e.g., mobile phones, cameras), edge servers have moderate, and cloud servers are the most powerful machines.
    \item \textbf{Network latency:} determines how efficiently entities involved in the \cn{} can communicate with each other, measured from data source to offload target.
    \item \textbf{Network type:} can be wired (Ethernet, USB) or wireless (WiFi, Bluetooth, cellular, WANs).
          The network type also determines what kind of mobility a developer can expect in the deployment.
    \item \textbf{Operator:} identifies the stakeholder running the infrastructures and development of workloads.
\end{enumerate}

\subsection{Key Characteristics with Existing Computing Models}
Having synthesized key characteristics, we now focus on existing computing models and how they consider the aforementioned characteristics.
We study 17 unique computing models in our survey that leverage a combination of cloud, edge, and endpoint devices, and enable workload offloading
between these devices.
Based on the amount of overlap (if the model has been superseded or merged), and the current interest (is part of active academic/industrial interest), we narrow down the discussion to five specific models while giving a rationale for the process.

\noindent\textbf{1. Mist Computing}
is concerned with forming a peer-to-peer (P2P) network of user-operated endpoint devices, where resource-constrained and/or busy endpoints can offload workload to nearby endpoints with more aggregate processing power or storage capacity~\cite{preden2015benefits}.
Mist computing is a representative of a larger group of P2P computing models for endpoint devices such as mobile crowdsourcing~\cite{ren2015exploiting}, peer-to-peer computing~\cite{milojicic2002peer}, mobile crowd computing~\cite{fernando2012honeybee}, mobile ad hoc cloud computing~\cite{drolia2013mobile}, and transparent computing~\cite{ren2017serving}, with each having its assumption on how the P2P network is formed, the goal of the offloading, and possible interaction with a centralized cloud or edge component.

\noindent\textbf{2. Edge Computing}
enables the offloading of computation and related data from endpoints to a cluster of edge devices through low-latency networks to enhance previous endpoint-only applications.
For applications that share data and state between tenants, like federated learning and online gaming, data can also flow from cloud to edge~\cite{DBLP:journals/jsac/WangTSLMHC19}.
Related to edge computing is edge-centric computing, which focuses on applications with human interaction.

\noindent\textbf{3. Multi-access Edge Computing:}
Previously known as mobile edge computing~\cite{etsi2016mobile}, MEC is concerned with augmenting cellular and wireless infrastructure (4G, 5G) with computing capacity to process user data from multiple endpoints close to users instead of in the cloud~\cite{taleb2017multi,etsi2022multi}.
Contrary to edge computing, where edge devices are owned by users or cloud providers, edge devices in MEC are exclusively owned by network providers and standardized under the European Telecommunications Standards Institute (ETSI).

\noindent\textbf{4. Fog Computing}
extends cloud services to edge to accelerate former cloud-only applications by leveraging the edge's low latency to endpoints~\cite{shi2016edge,aws2021edge}.
Fog computing leverages multiple hierarchical edge clusters: From resource-constrained devices near users to micro data centers near the cloud, providing a trade-off between compute capacity and network latency.
Offloading services to micro data centers is the focus of cloudlet computing, a precursor of fog computing~\cite{satyanarayanan2009case}.

\noindent\textbf{5. Mobile Cloud Computing:}
Endpoints offload compute tasks and related data for workloads without strict latency constraints directly to the cloud to benefit from the cloud's cheaper and more stable resources~\cite{dinh2013survey}.
Even selected latency-constrained applications can directly offload to the cloud as the average latency between cloud and endpoint is decreasing due to better networking infrastructure and more cloud data centers~\cite{corneo2021surrounded}.
Mobile grid computing is a precursor of mobile cloud computing, leveraging grids instead of clouds~\cite{guan2005grid}.
In situ computing explicitly focuses on preprocessing data on endpoint devices before offloading to reduce network load~\cite{li2015insitu}.
Osmotic computing combines mobile cloud computing and edge computing, advocating for a dynamic and seamless offloading process to cloud or edge~\cite{villari2016osmotic,satyanarayanan2017emergence}.

\subsection{A Case for the Unification}
In the past sections, we have synthesized key characteristics (\textit{what a developer should know}) and identified how existing computing models provide (or lack) guidelines for these key characteristics for \cc{} workloads.
Based on this discussion, we make a case for the unification of different computing models based on three primary arguments.
First, finding a suitable model now requires comparing requirements from a particular workload to the features provided by the five models (Table~\ref{tab:computing}) and selecting the best fit.
This approach of considering each model in isolation offers a restricted view of the task offloading design space.
For example, data processing applications may require computing services in the edge and storage services in the cloud~\cite{vasisht2017farmbeats}: Edge computing offers the former, and mobile cloud computing offers the latter, but no computing model explicitly offers these services combined.
This shows that choosing an existing computing model with fixed characteristics for a particular workload deployment has major limitations that create \textit{silos} of knowledge, and hence, confusion for users.
Second, computing models often only focus on compute/task management-related responsibilities.
However, in practice, data and runtime state management and resource allocation with multi-workload/tenant situations are equally important~\cite{2020-hotedge-griffin}.
For example, mist computing gives guidelines regarding how to establish and push computation in a P2P network of endpoint devices but provides little guidance on how data sharing is done.
Lastly, providing end-to-end data provenance and privacy properties is challenging without building an end-to-end view of the workload.
For example, in multi-access edge computing, where network providers own a part of the infrastructure and resources (not the workload developers), it would necessitate cooperation with the providers to ensure a secure resource allocation and isolated execution, a property challenging even in cloud computing~\cite{2021-middleware-delft-isolation,2020-hotcloud-stratus}.

To close the knowledge gaps, in this work, we propose to take an \textit{ab initio} approach for designing a unified \cn{} computing model and an associated unified reference architecture in the next section.
\section{The \org{} \CC{}\\\RA{}} \label{sec:refarch}
\begin{figure}[t]
    \centering
    \includegraphics[width=.95\linewidth]{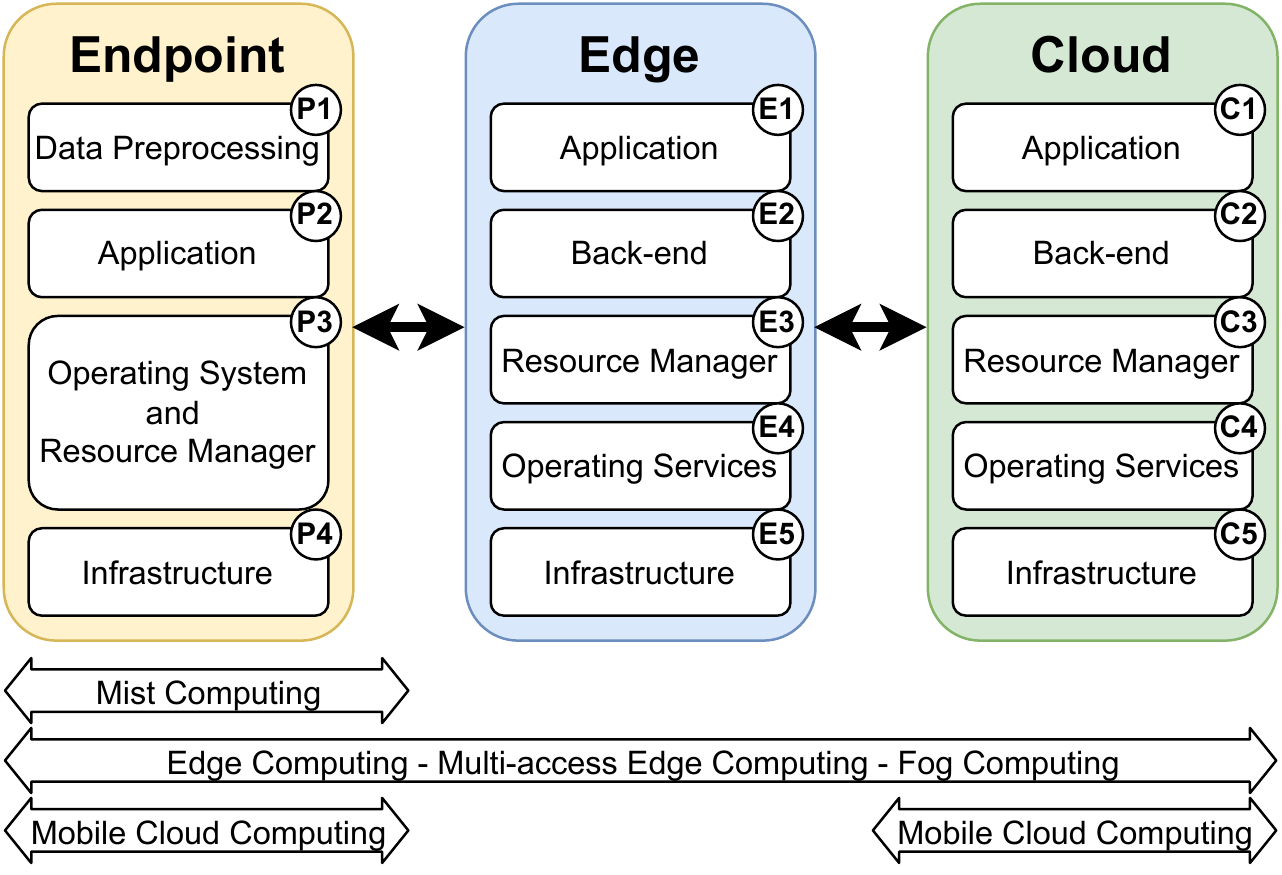}
    \caption{Reference architecture for the \cc{}. The computing models are mapped to the parts of the architecture relevant to them.}
    \label{fig:refarch}
    \vspace{-0.4cm}
\end{figure}

Having made a case for a unified computing model, we present our unified reference architecture for the \cc{} in Figure~\ref{fig:refarch}.
A computing model typically has a reference architecture that describes the components a computing model operates on and the interactions between these components.
All five selected models have reference architectures, with the most prominent architectures listed in \Sec{sec:relatedwork}.
The goal of building a unified reference architecture is to abstract away the specifications of the underlying hardware and peculiarities of specific workloads and discuss more broadly the foundational building blocks of the \cc{} regarding compute, data, and resource management.

\subsection{Overview and Design Process}

\noindent\textbf{Overview:}
The \org{} \ra{} consists of three tiers of systems: cloud, edge, and endpoint, with their associated components and responsibilities, as shown in Figure~\ref{fig:refarch}.
The three-tier system is taken from the developer's view of compute offloading and data processing: endpoint devices are the last connected components in the architecture that create data streams for long-term storage and processing in cloud data centers, located at the top of the architecture, or require data hosted in the cloud for content delivery networks~\cite{2019-hotedge-cnd-p2p2} and applications with shared state~\cite{DBLP:journals/jsac/WangTSLMHC19}.
In between, the edge represents multiple hops of processing and storage capabilities.
In explaining the \ra{}, we will borrow well-defined cloud-centric components and extend them to include the expanded \cn.
Lastly, we show the utility of the proposed \ra{} by \textit{mapping} it into two concrete \cn{} workloads: machine learning and industrial IoT in \Sec{sec:mapping}.

\noindent\textbf{Design process:}
The primary requirement from a reference architecture is that it should capture various deployment models that we discussed in the previous section.
We start by analyzing and identifying commonalities and differences between the five selected computing models and use the overlap between the models as the basis of the architecture, with the differences between the models becoming implementation details of the components~\cite{mouradian2018comprehensive}.
For example, all selected computing models (some explicitly, others implicitly) use resource managers to manage the complex stack of cloud, edge, and endpoint resources and the networks connecting them, but the specific implementation of these resource managers differs widely between workload deployments.
As such, resource managers should be a core component of the architecture, while the implementation of the resource manager should be an implementation detail of the component.
By creating a unified architecture for the edge continuum, we present a single platform for research into cloud, edge, and endpoint resources, unifying previous research efforts and allowing exploration in new research directions.
We are the first to create such a unified end-to-end architecture, extending previous work which looked at computing models in isolation (\Sec{sec:relatedwork}).

In the next sections, we discuss (i) the design of the architecture's components; (ii) the rationale behind their position in the \ra{}; and (iii) how to offer compute, data, and resource management services.

\subsection{Endpoint Components}
Endpoints are the last hop of processing and connectivity to users with the following key features: First, they often are data sources, e.g., user input through online gaming~\cite{zhang2019improving}, environment sensing~\cite{vasisht2017farmbeats}, or video surveillance~\cite{khan2019deep}.
Second, they are resource and energy-constrained, e.g., IoT devices such as sensors, and embedded systems~\cite{satyanarayanan2009case}.
Lastly, they may be mobile, e.g., smartphones, where location-sensitive services must migrate to the device's location~\cite{2017-sec-mmog-mobility}.
To support the unified computing model, we have split up endpoint responsibilities into the following four components:

\noindent \textbf{P1. Data Preprocessing:} 
As data generators, we identify endpoints to be perfectly positioned to run early workload-specific data preprocessing logic such as filtering, tracking, aggregation, etc.
This preprocessing aims to reduce the amount of data pushed to edge and cloud servers. 
Preprocessing capabilities can be built-in as a hardware accelerator or a software component available for use in high-level applications, thus freeing them to do low-level data preparations.
For example, in video surveillance, camera endpoints can support built-in object tracking or face detection~\cite{hu2015case}, as can be the case for self-driving vehicles~\cite{lin2018architectural}.
The key decisions here for developers are: (1) does data need preprocessing; (2) can it be preprocessed at the endpoint itself, or must it be offloaded; (3) when and how to offload compute/data for preprocessing and collect results; and (4) can preprocessing be helped with hardware acceleration for energy efficiency.

\noindent \textbf{P2. Application:} 
Applications contain user-defined logic that runs on the endpoint to process incoming data (from edge/cloud or post-preprocessing stage) and make decisions.
Typical logic in the application can be running heuristics, making offloading decisions, monitoring performance metrics, and/or triggering changes 
in the application deployments.
For example, an application can join a P2P network with other endpoints to do mist computing or display model data rendered by cloud servers in an online gaming setting (data flow from cloud to endpoint).
The key decisions here for developers are: (1) in which direction compute and data offloading happens - P2P or hierarchically to edge/cloud; (2) when and how to offload compute/data for processing and collecting results; (3) what are favorable conditions for offloading and how to quantify them; and (4) how to reserve and allocate local (endpoint) and further (edge, cloud) resources.

\noindent \textbf{P3. Operating System and Resource Manager:} 
In contrast to edge/cloud servers, endpoints are typically single-tenant~\cite{satyanarayanan2019computing}; hence, OS-level resource managers (RMs) and multiplexers are often enough to meet the workload needs. 
Common examples of OS-level RMs are Android, iOS, TinyOS, and QNX~\cite{hill2000system}.
These RMs are optimized for deployed hardware and scenarios such as the presence or absence of specific hardware features (virtualization, secure enclaves), energy management, high priority for user interactions, etc.
The key decisions here for developers are: (1) what kind of resources does a workload need - CPU, memory, hardware accelerators, networking capacity; (2) how to reserve/allocate resources using the OS-level RMs; and (3) do RMs meet the functional (performance SLAs) and non-functional (cost, fault handling, data corruption) requirements of the workload?

\noindent \textbf{P4. Infrastructure:} 
Infrastructure includes all physical computing, memory, network, and storage resources available to the operating system. 
These resources can be general purpose (e.g., smartphones) or specialized for mobility, energy efficiency, etc. (e.g., IoT, embedded hardware).
Unlike cloud data centers, the infrastructure at the endpoint can be owned and operated by application users and developers besides cloud providers.~\cite{aws2021edge}.
The key decisions here for developers/providers are: (1) what are the most popular workloads, computing models, and deployment modes leveraging the endpoints; (2) is there sufficient raw network, storage, and processing capabilities available; (3) what is the appropriate cost and scaling model (horizontal or vertical scaling); and (4) how to integrate and manage infrastructure in a unified \cn{}~\cite{2018-hotedge-openstack}.

\subsection{Edge Components} \label{sec:edge_node}
The \ra{} in Figure~\ref{fig:refarch} shows that edge and cloud share the same high-level design.
This similarity is because both can run multi-tenant workloads on shared infrastructure and so need to offer similar services.
However, uniquely at the edge, there should be support for application offloading both vertically (cloud to edge and back) and horizontally (from one edge system to another)~\cite{mortazavi2017cloudpath}.
The presence and absence of such capabilities determine what computing model one can use at the edge, e.g., edge clusters can not communicate among themselves with edge computing, while they can with fog computing.
These unique capabilities are implemented in the edge components.
To encompass these possibilities, we define these components for edge systems:

\noindent \textbf{E1. Applications} 
are at the first step from the endpoints, and hence they are in the best position to make decisions regarding placements, offloading, and scheduling of application components to meet workload-specific objectives.
For example, applications can either do processing at the edge or preprocess data and offload more complex workloads to the cloud~\cite{kang2017neurosurgeon}.
The key decisions here for developers are: (1) in which direction does data and compute offloading happen (cloud to edge, or vice versa); (2) what are the location, data movement, and mobility-related requirements; and (3) are back-end and operating services provided services enough to meet the workload demands.

\noindent \textbf{E2. Back-end:} 
Represents more general-purpose application execution frameworks such as TensorFlow Lite and WebAssembly runtime~\cite{gadepalli2020sledge}.
These back-end frameworks typically can manage application-specific memory, storage, communication, and workflows.
The key decisions here for developers are: (1) does the back-end supports the desired offloading model; and (2) what kind of computing, storage, and networking resources does a back-end need.
    
\noindent \textbf{E3. Resource Manager:} 
Manages an edge system's application-independent \textit{physical and virtual resources} (such as virtual machines and containers).
These resource managers can be local or distributed, and their architecture determines what computing model one can run on the infrastructure.
For example, KubeEdge is a hierarchical edge resource manager for Kubernetes, which does two levels of independent allocation for cloud- and edge-related resources.
The key decisions here for developers/providers are: (1) what execution environment does an application want; (2) what are the scalability requirements; (3) what are the environment allocation overheads; and (4) what are data privacy and security level restriction in the scheduling of resources.
    
\noindent \textbf{E4. Operating Services:} 
Provides support to build distributed applications, and their responsibilities include (but are not limited to) communication~\cite{light2017mosquitto}, metadata management~\cite{erwin2021edge}, consensus services~\cite{hao2018edgecons}, monitoring~\cite{superedge2021}, storage services~\cite{gupta2018fogstore}, etc.
The key decisions here for developers/providers are: (1) is there an operating service for different kinds of communication and coordination patterns; (2) are services protected and isolated from tenants; (3) how do services handle resource allocation for themselves; and (4) what are scalability, performance, and fault-tolerance requirements for such services.

\noindent \textbf{E5. Infrastructure:} 
Similar to endpoints, compute, memory, and storage resources are provided to the resource manager.
However, unlike endpoints, resources are split into physical and virtual resources.
Bare-metal deployments give users direct access to hardware, while virtualization technologies like virtual machines and containers abstract physical resources away to provide more flexibility and security for a slight performance penalty.
The key decisions for developers/providers are: (1) how to split (or multiplex) software and hardware resources securely between tenants; and (2) is there enough cooling, energy, and infrastructure to meet the demands.

\subsection{Cloud Components}
We identify two critical roles for cloud in the \cc{}: The first is that of a central controller that manages multiple \cc{} resources and schedules workload on them~\cite{wang2017enorm,superedge2021}.
The controller has a global overview of the endpoint, edge, and cloud resources.
Developers upload \cn{} workloads to the cloud, which the cloud controller can schedule on specific edge systems.
The global overview allows the central controller to schedule work more efficiently than a decentralized approach where each edge system manages itself.
However, maintaining a global overview and scheduling workload from the cloud adds extra overhead to the offloading process~\cite{xiong2018extend}.
The second role is that of an offloading target that uses cloud computing to provide much more computing and storage capacity than edge~\cite{microsoft2021edge}.
Resources in the edge are limited~\cite{ecc2017edge}, so cloud computing can be used to run resource-intensive applications like deep learning instead.
However, applications that require low end-to-end latency can not leverage the cloud due to the increased communication latency compared to the edge~\cite{linux2021state}.
The role of offloading target in the \cn{} is key to both edge and cloud; therefore, they share the same high-level design in our architecture. 
The components from edge, described in \Sec{sec:edge_node}, can also be applied to cloud.
Liu et al.\/ already provide a reference architecture for cloud computing~\cite{liu2011nist}.

\subsection{Domain-specific Architectures} \label{sec:mapping}

\begin{figure}[t]
    \centering
    \includegraphics[width=.9\linewidth]{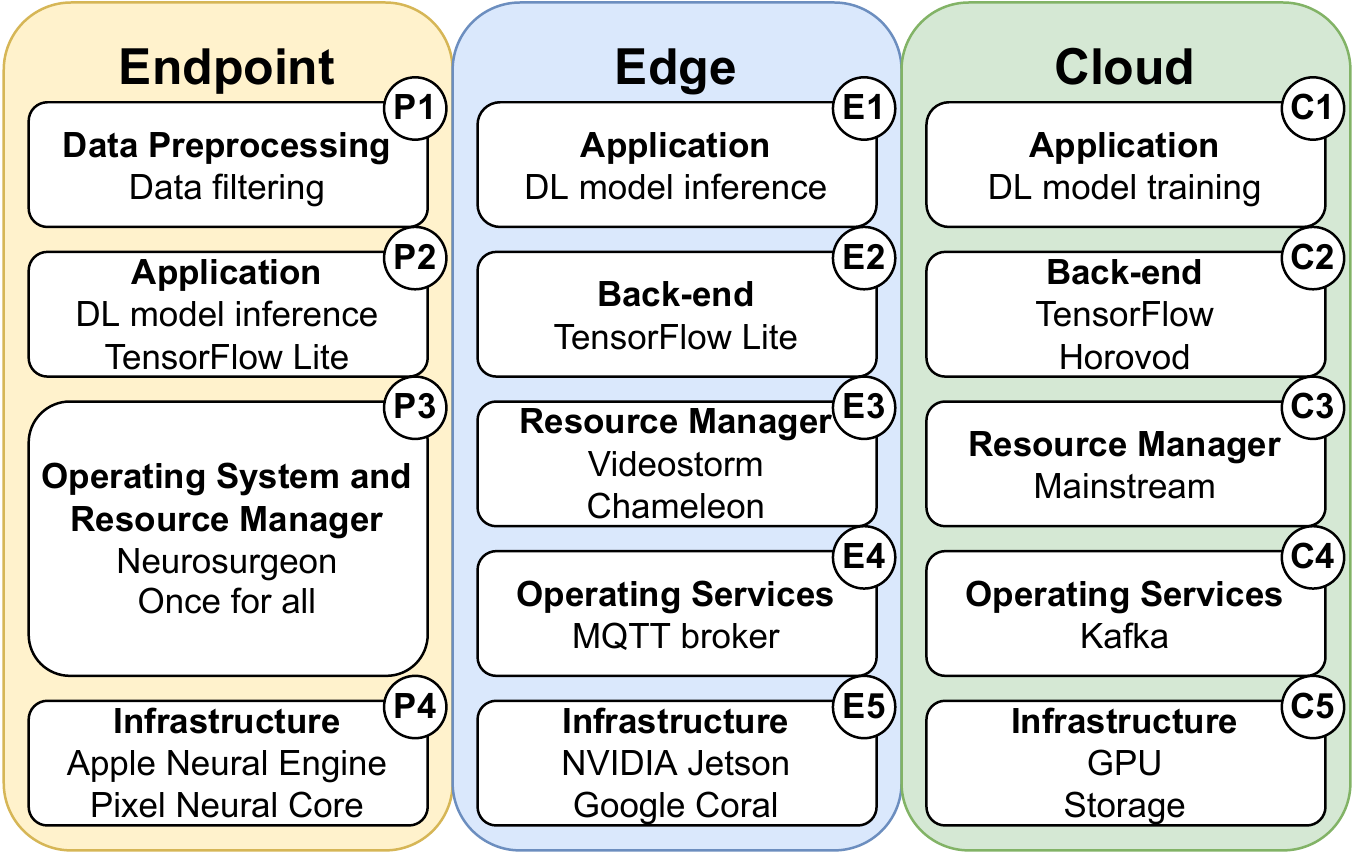}
    \caption{Deep learning architecture. Examples include Neurosurgeon~\cite{kang2017neurosurgeon}, Once for all~\cite{https://doi.org/10.48550/arxiv.1908.09791}, Videostorm~\cite{DBLP:conf/nsdi/ZhangABPBF17}, Chameleon~\cite{DBLP:conf/sigcomm/JiangABSS18}, and Mainstream~\cite{DBLP:conf/usenix/JiangWCTMKKPAG18}}
    \label{fig:ml-mapping}
    \vspace{-0.5cm}
\end{figure}

\begin{figure}[t]
    \centering
    \includegraphics[width=.9\linewidth]{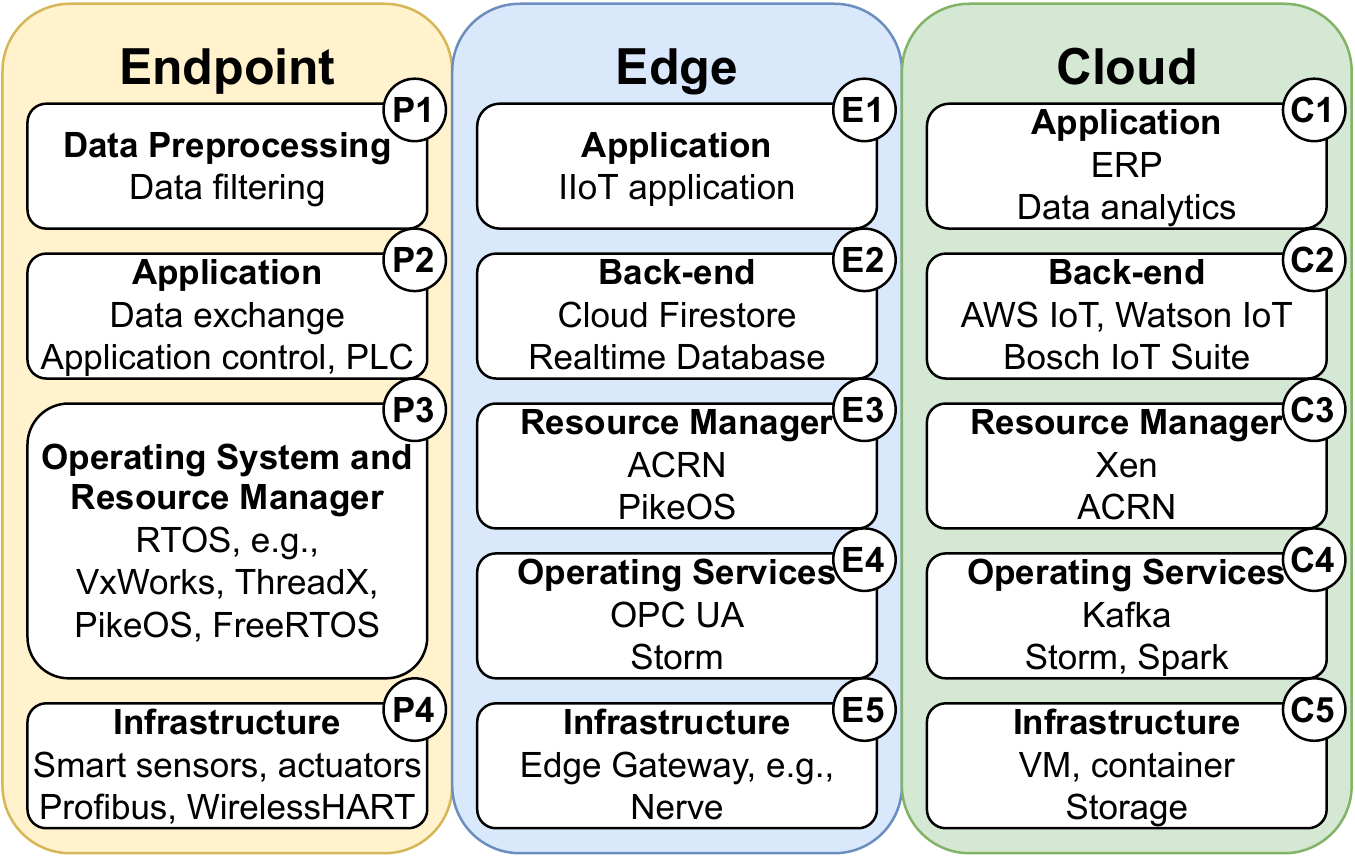}
    \caption{Industrial IoT (IIoT) architecture. Examples include Cloud Firestore~\cite{google2021databases}, ACRN~\cite{arcm2021hypervisor}, and PikeOS~\cite{pikeos2021hypervisor}.}
    \label{fig:iiot-mapping}
    \vspace{-0.5cm}
\end{figure}

We create architectures for deep learning (Figure~\ref{fig:ml-mapping}) and industrial IoT (IIoT, Figure~\ref{fig:iiot-mapping}) use cases and provide example systems for each component to demonstrate how the unified \ra{} can be used to explore design trade-offs for application deployments in the \cc{}.

\textbf{Deep Learning:}
An important trend for deep learning in the \cn{} is that model training and inference tasks are split across multiple tiers of devices, as deep learning applications train neural networks with large memory footprints on data generated on endpoints.
Deciding where and how to deploy these tasks presents complex trade-offs and requires careful analysis~\cite{kang2017neurosurgeon}.
For example, cloud services like AWS SageMaker (component C2 in Figure~\ref{fig:ml-mapping}) offer high-performance managed machine learning training in the cloud but require possibly privacy-sensitive data to be moved to public infrastructure.
Federated learning provides a solution as users offload anonymized information of their privately trained model to the cloud instead of the data used to train the model.

While model inference is much less compute and storage intensive than model training, it still requires the use of specialized deep learning frameworks like Neurosurgeon~\cite{kang2017neurosurgeon} (P3) and models like MobileNet~\cite{sandler2018mobilenet} to be deployable on constrained edge or endpoint devices.
These specialized frameworks and models present a complex trade-off between response time and model accuracy: By lowering the compute and storage requirements for model inference, model accuracy drops, but the application can be deployed on constrained devices close to the user, lowering the response time to the user.
For recommender systems, real-time user feedback may be required, so response time is preferred above model accuracy, while for video analytics the opposite may apply.

\begin{table*}[t]
    \centering
    \caption{Selection of Parameters Offered by the Framework.}
    \label{tab:api}
    \begin{tabular}{l|l|l}
    \specialrule{.1em}{.0em}{.0em} 
    Parameter                   & Architecture Component    & Description                                                                           \\
    \specialrule{.1em}{.0em}{.0em}
    Data generation frequency   & P1                        & Rate at which data is generated at endpoints.                                         \\
    Application                 & P1, P2, E1, E2, C1, C2    & Application to deploy and benchmark.                                                  \\
    Resource manager            & P3, E3, E4, C3, C4        & Resource manager to deploy in the continuum.                                          \\
    Hypervisor                  & P4, E5, C5                & Virtual machine provider, e.g., QEMU.                                                 \\
    Devices per tier            & P4, E5, C5                & Number of cloud, edge, and endpoint devices to emulate.                               \\
    Cores per device            & P4, E5, C5                & Number of CPU cores to assign to each VM.                                             \\
    Quota per CPU               & P4, E5, C5                & Use part of a CPU core to emulate slower hardware.                                    \\
    Network per tier            & P4, E5, C5                & Throughput and latency between emulated devices.                                      \\
    Machine addresses           & P4, E5, C5                & Machine IP addresses when doing emulation across multiple physical machines.          \\
    \specialrule{.1em}{.0em}{.0em} 
\end{tabular}

\end{table*}

\textbf{Industrial IoT:}
In IIoT, endpoint devices generate large amounts of data that often need to be processed in real-time, with strict requirements for privacy and durability~\cite{algumaei2018etfa}.
These endpoints are connected to Programmable Logic Controllers (PLCs, component P2 in Figure~\ref{fig:iiot-mapping}); these are control systems for local control without support for advanced processing due to resource limitations.
Thus, offloading to remote devices over a fieldbus or via wireless communication is required (P3).

As many industry deployments include a vast amount of sensors and actuators, processing and storing large amounts of generated data requires extensive resources. 
Therefore, public cloud offerings like AWS IoT and Bosch IoT suite (C2) can be a good fit for many deployments~\cite{algumaei2018etfa}.
However, reliable real-time processing guarantees may be broken if the network infrastructure between the endpoints and cloud can not support the large data streams.
Moreover, offloading sensitive data to third parties may introduce security and privacy concerns.
On-premise cloud devices are therefore often used in IIoT deployments, delivering more stable performance by eliminating possible connectivity issues to remote clouds and guaranteeing data privacy.
In addition to clouds, edge gateways (E5) can be used to offer prompt response time and increased security.

\textbf{In conclusion}, with the example architectures for deep learning and IIoT we show that our uniform architecture helps developers and infrastructure providers navigate the \cc{} without being restricted to hardware and software solutions from a single computing model, but can freely combine solutions instead.

\section{Deployment and Benchmarking Framework} \label{sec:results}
In the design of our reference architecture, we showed that by changing the implementation of components in our architecture, we can create deployments for any task offloading computing model.
To demonstrate this mechanic in practice, we present Continuum, an infrastructure deployment and benchmarking framework for the \cc{} in this section and an accompanying analytical model in \Sec{sec:perf_model}.
The combination between infrastructure deployment and benchmarking, both highly configurable through a list of parameters, allows us to perform a systematic quantitative exploration of the \cc{} deployment space by performing performance analysis on architecture components.
We attempt to answer three fundamental questions here:

\begin{enumerate}
    \item \textit{First, does our framework allow exploring different complex deployments in the \cc{}?}
          We explore different computing models in the framework using a few lines of changes in the configuration setup with parameters derived directly from the \ra{} (Table~\ref{tab:api} and example Listing~\ref{lst:config})).
    \item \textit{Secondly, how can our framework help explore design trade-offs?}
          Using the framework, we provide a breakdown of end-to-end latency (compute, communication) with an aggregation factor for various cloud, edge, and endpoint offload targets (Figures~\ref{fig:bench-large-breakdown} and ~\ref{fig:bench-endpoint}).
    \item \textit{Lastly, does the analytical model provide exploratory guidelines in line with the empirical evaluation?}
          We explain our first-order analytical model, corroborate its predictions with empirical evaluations, and show its offloading guidelines as a heatmap (Figure~\ref{fig:heatmap}).
\end{enumerate}

The framework and model are open-sourced and available at \url{https://github.com/atlarge-research/continuum}.

\subsection{Framework Design and Interaction}\label{sec:eval-design}
\begin{figure}[t]
    \centering
    \includegraphics[width=0.9\linewidth]{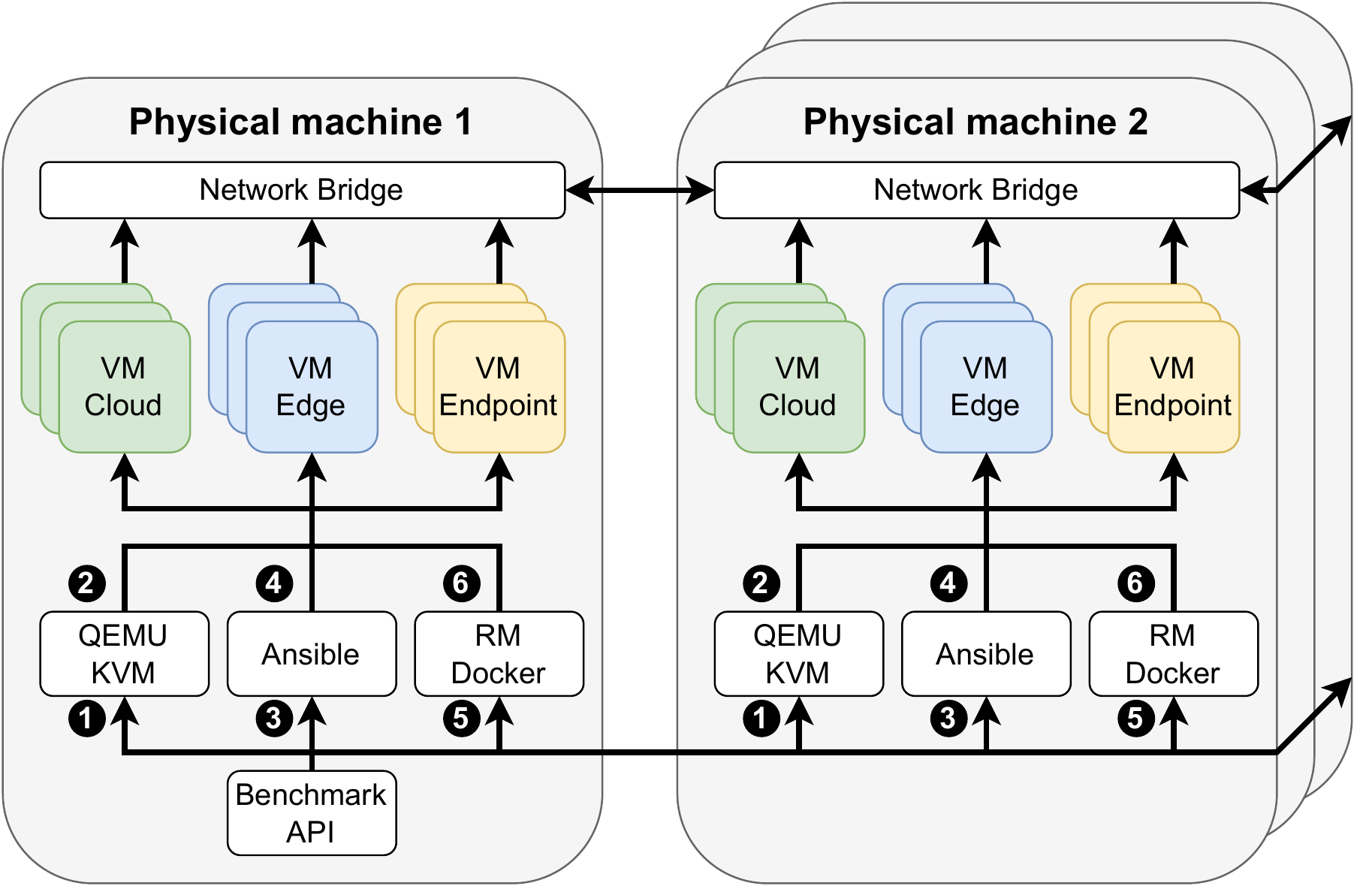}
    \caption{Design of our \cn{} benchmarking framework.}
    \label{fig:bench_design}
\end{figure}

Performing experiments in the \cn{} is challenging due to the wide range of networks and hardware available (Figure~\ref{fig:sys-model}).
Currently, no physical infrastructure is available that allows the exploration of all task offloading computing models and related deployments, only a selection~\cite{csenel2021edgenet}.
As an alternative, devices and networks can be virtualized to emulate the \cc{} on commodity hardware.
This emulation offers a flexible and easy-to-use method to explore the \cn{}'s deployment space, as emulated hardware is configurable, unlike physical hardware.
Furthermore, such an emulated environment can be used to benchmark various deployment scenarios using components from our reference architecture.
We have created such an emulated deployment and benchmarking framework and present its design in Figure~\ref{fig:bench_design}.
Our framework allows the emulation of many virtual machines across many physical machines, ranging from commodity hardware to specialized edge or endpoint hardware.
This flexibility allows our framework to function as an emulated benchmarking environment on general-purpose hardware and as a benchmark on physical \cc{} resources.
The framework offers a set of simple and intuitive parameters that can be explored and mapped to the \ra{} to understand their implications on workload performance, as shown in Table~\ref{tab:api} and abridged in Listing~\ref{lst:config}.

On the infrastructural level (components P4, E5, and C5 in Figure~\ref{fig:refarch}), users can specify the number of required cloud, edge, and endpoint devices (line 6) that should be emulated, the specifications of the emulated devices (cores, CPU share/quota, memory, storage, lines: 7-16), and the networks connecting the devices (component \designref{1} in Figure~\ref{fig:bench_design}).
We currently leverage the Linux QEMU/KVM hypervisor~\cite{bellard2005qemu} (v$6.0$) with virtual machines (VMs) and tools (\texttt{tc} and \texttt{blkiotune}) to emulate a large \cn{} setup (\designref{2}) on a 3-machine Xeon Silver 4210R CPU cluster connected with a 1 Gbps link.
For the next step, operating services and resource managers are installed in the provisioned VMs (P3, E3, E4, C3, C4) through Ansible~\cite{hochstein2017ansible} (\designref{3} and \designref{4}).
We currently support Kubernetes (v$1.21$) as cloud resource manager, KubeEdge (v$1.8$) as edge resource manager, and OpenFaaS for serverless resources.
On the application level (P1, E1, C1), we support data-processing applications with data generation at the endpoints and processing options at an endpoint, edge, and cloud (\designref{5} and \designref{6}).

\begin{lstlisting}[language=Ini,basicstyle=\ttfamily\footnotesize,frame=single,
caption={Example configuration file for Figures~\ref{fig:bench-large-breakdown} and ~\ref{fig:bench-endpoint}.},captionpos=b,float=tp,belowskip=-1em,label={lst:config}]
[infrastructure]
hypervisor = qemu
thread_pinning = True

# VM settings for cloud, edge, endpoint
devices_per_tier = 10,0,40
cores_per_device = 4,0,1
quota_per_cpu = 1.0,0,0.5

# Latency (ms): average,variability
cloud_to_cloud = 1,0
cloud_to_endpoint = 45,5

# Throughput (Mbit): average
cloud_to_cloud = 1000
cloud_to_endpoint = 8

machine_address = 192.168.1.1,192.168.1.2

[benchmark]
use_benchmark = True
data_generation_frequency = 5
application = image_classification
resource_manager = kubernetes
\end{lstlisting}

\subsection{Offload Model Design Space Exploration}
This section illustrates how a developer explores the \cn{} design space using the aforementioned simple configurations with an image-processing workload as an example.
In this workload, images are generated at endpoints (e.g., camera, thermal sensors), and can be processed at endpoints, edge, or cloud nodes with different CPU processing capabilities.
We fix the image generation rate to five images per second, and set the bandwidth between the endpoint and its offloading targets to 8 Mbit/s, a representative throughput value for 4G networks~\cite{DBLP:conf/mmsys/RacaQZS18}.

\begin{table}[t]
    \centering
    \caption{Parameters for the Deployments Used in Our Evaluation.}
    \label{tab:parameters}
    \adjustbox{width=\linewidth}{
        \begin{tabular}{l|llll}
    \specialrule{.1em}{.0em}{.0em} 
    Parameter                   & Cloud         & Edge-Large    & Edge-Small    & Mist      \\
    \specialrule{.1em}{.0em}{.0em}
    Resource manager            & Kubernetes    & KubeEdge      & KubeEdge      & -         \\
    Worker location             & Cloud         & Edge          & Edge          & Endpoint  \\
    Workers                     & 10            & 10            & 10            & 10        \\
    Worker cores                & 4             & 4             & 2             & 2         \\
    Worker quota                & 1.0           & 1.0           & 0.75          & 0.5       \\
    Endpoints per worker        & 4             & 4             & 2             & 1         \\
    Network latency (ms)        & 45            & 30            & 7.5           & 7.5       \\
    (\#cloud, \#edge, \#endpoint) & (11, 0, 40) & (1, 10, 40)   & (1, 10, 20)   & (0, 0, 20) \\
    \specialrule{.1em}{.0em}{.0em} 
\end{tabular}

    }
\end{table}

Table~\ref{tab:parameters} shows how varying different parameters in listing~\ref{lst:config} allows us to explore four computing offload models on our 3-machines cluster.
For example, with \textit{cloud} offloading, images are transmitted from the endpoint to a cloud cluster with 10 workers, each with 4 cores and full CPU quota (1.0), and each worker serves 4 endpoints (controlled via parameters \texttt{devices\_per\_tier, cores\_per\_device, quota\_per\_cpu}).
There are 11 cloud nodes in total (10 workers and one controller), zero edge nodes, and 40 endpoints.
In contrast, with \textit{edge offloading}, workers can be placed at the edge using the parameters \texttt{devices\_per\_tier} and \texttt{quota\_per\_cpu} for large and small CPU configurations.
Lastly, to capture \textit{mist computing}, all image processing tasks are offloaded to other endpoint devices, thus marking the cloud and edge devices as zero (setting \texttt{devices\_per\_tier} as \texttt{0, 0, 40}).
The \Cn{} framework also allows us to specify the network and storage properties.

\begin{figure}[t]
    \centering
    \includegraphics[width=0.9\linewidth]{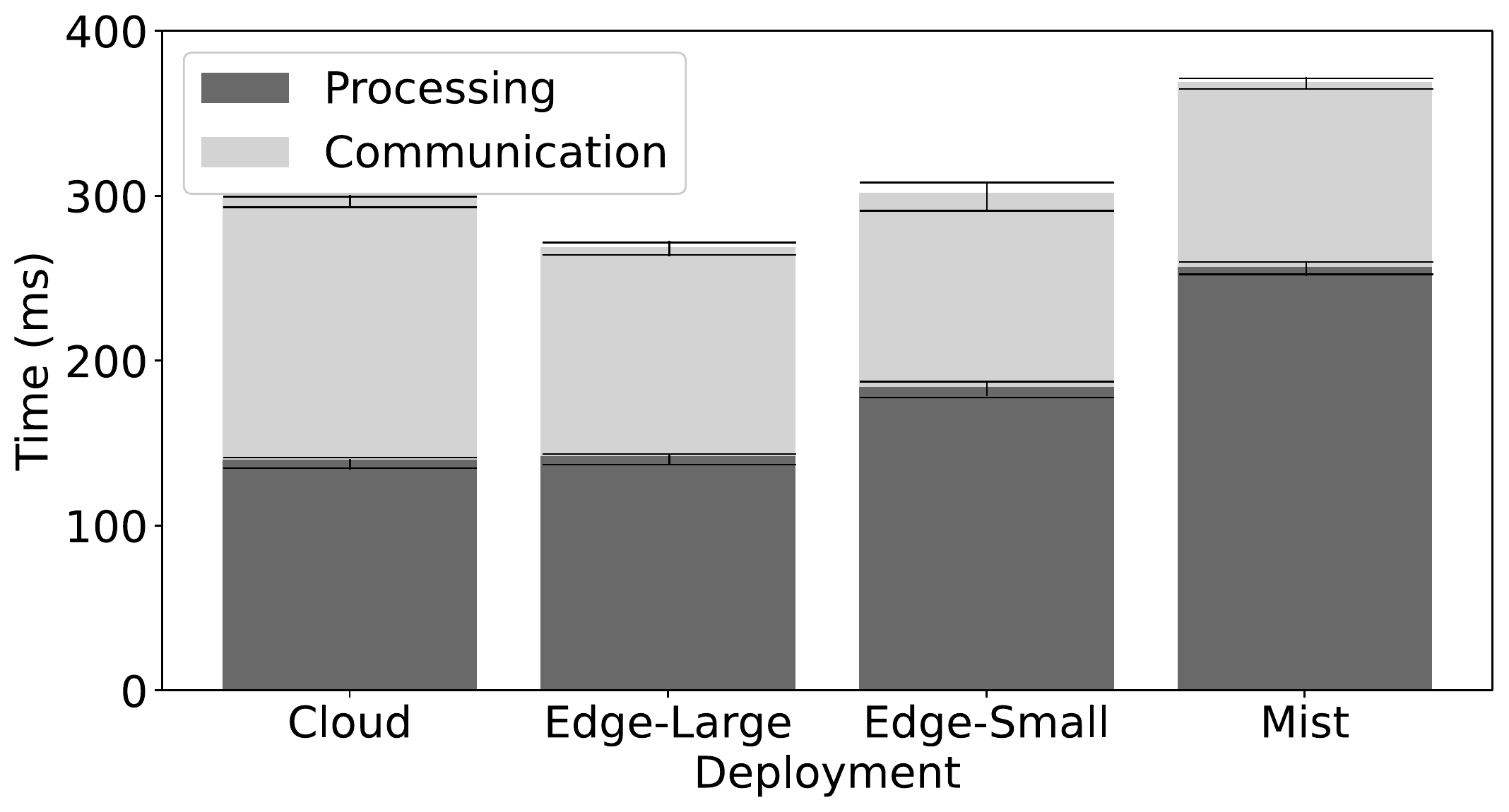}
    \caption{Breakdown of the end-to-end latency per deployment.}
    \label{fig:bench-large-breakdown}
\end{figure}

For these four computing models, figure~\ref{fig:bench-large-breakdown} shows our results.
The x-axis shows the offloading model.
The y-axis shows the end-to-end image processing time (in milliseconds, lower is better) split between computation and communication components.
We report the average and standard deviation values over three runs.
There are three main observations here.
First, our framework allows explorations of these different offloading models with less than 5 lines of configuration changes between them (total size 50 lines, not shown).
Second, we see a latency reduction between Cloud and Edge-large devices with equal CPU capabilities (the first two bars from the left) due to a decrease in communication time, as edge nodes are closer to the endpoint than cloud nodes.
Lastly, as we move closer to the data source (Edge-small and Mist), we receive a gradual decrease in computing capabilities (0.75 and 0.5 CPU fractions) while still reducing communication latencies.
However, gains from the latency reduction can not offset the overheads in computing due to slower processors; hence, the end-to-end processing latencies increase.
Depending on the workload requirements, a developer can decide which model is of interest to them.
For example, if the cutoff latency is 300ms, then a user must only consider Cloud or Edge-large, not Edge-small and Mist offloads.

\begin{figure}[t!]
    \centering
    \includegraphics[width=\linewidth,height=4.8cm]{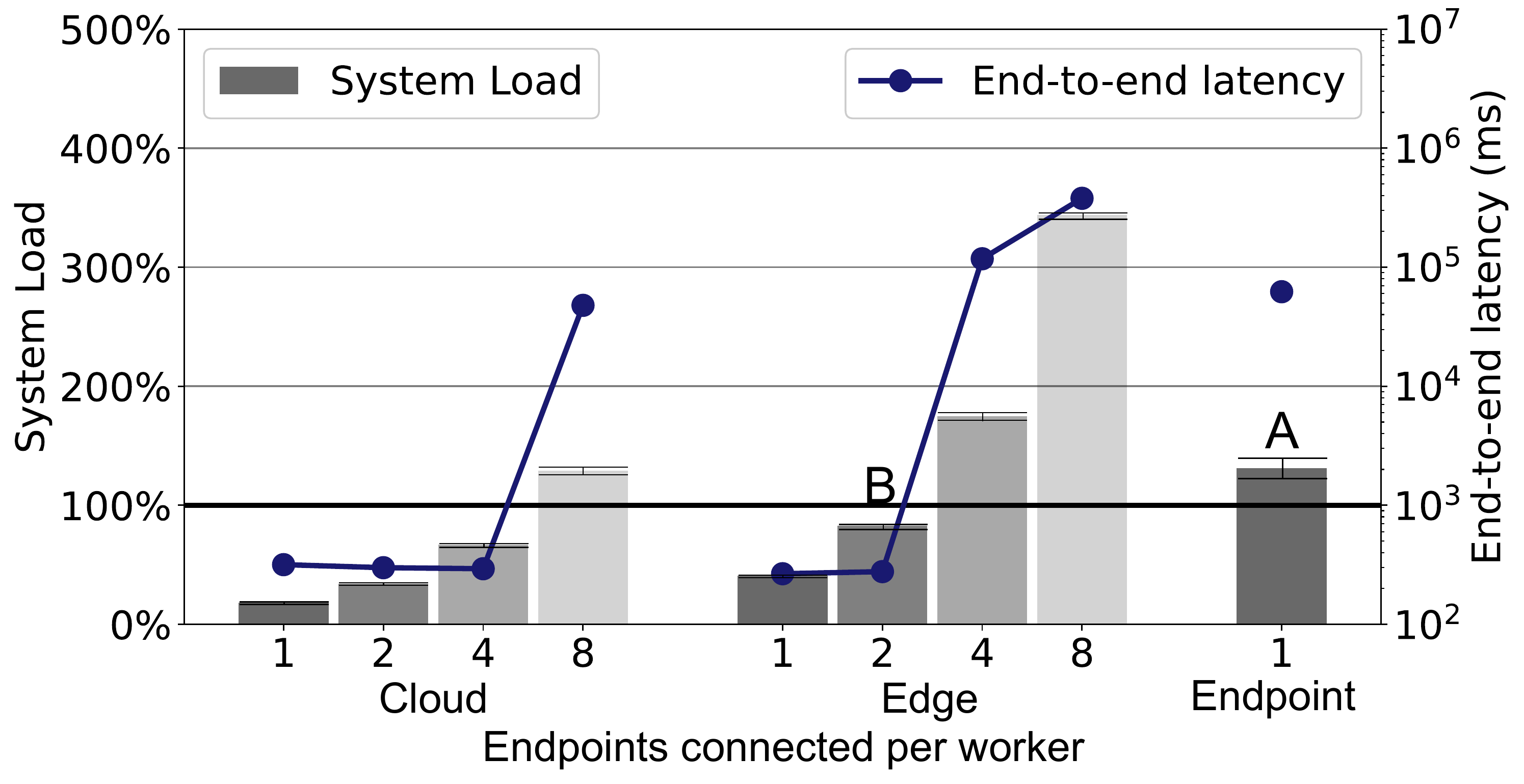}
    \caption{System load when processing data, with more endpoints connected to one processing device.
        Experiments A and B are used for the performance model examples in Equations~\ref{eqn:local_fill} and \ref{eqn:offload_fill}, respectively.}
    \label{fig:bench-endpoint}
\end{figure}

In the second experiment, we explore the cardinality and location of data aggregation (P1, E1, and C1 boxes) by varying how many endpoints an offload target can serve.
We identify this limit by defining a ``system load'' property, which is calculated by comparing the number of images processed (\textit{compute capacity}) to images offloaded (\textit{compute demand}) per second.
A load of more than 100\% implies more offloading requests are coming in than the worker device can process, resulting in queuing delays.
Such setups are common where a developer needs to identify the best place for data aggregation to balance out computation and communication capabilities in the \cn{}.
Close-to-source aggregation at endpoints reduces network transmission size but at the expense of using slower processors.
We benchmark a single offload task with an aggregation cardinality of 1, 2, 4, and 8 connected endpoints.
Figure~\ref{fig:bench-endpoint} shows our results.
The x-axis shows the aggregation granularity (cardinality and location), the left y-axis (the bar) shows the system load, and the right y-axis (log-scale, not starting from 1) shows the end-to-end latencies in milliseconds.
Looking at the systems loads, we can observe that cloud-based aggregation can support up to 4 endpoints/task, while edge only supports 2 endpoints/task, after which the system load increases beyond 100\% because a queueing delay of image processing requests is introduced in the end-to-end latencies.
The graph also shows an endpoint-only bar that represents a locally-processed solution that is non-viable as the system load is more than 100\%, thus failing to deliver real-time processing because of queueing delays.

\textbf{The key takeaway messages} with these experiments are that the open-source \Cn{} framework allows (i) a \textit{comprehensive exploration} of various compute offloading models with an expressive list of parameters; and (ii) \textit{a deep exploration} of specific setups such as aggregation point offloading and cardinality.
\section{Analytical Performance Model} \label{sec:perf_model}
For our final contribution, we enhance the performance analysis capabilities of our \cc{} benchmark by introducing a simple-to-use, first-order analytical performance model similar to a Roofline model in Figure~\ref{fig:heatmap}.
The model predicts if applications can be offloaded to cloud or edge, should be processed locally on endpoints, or are not suited for deployment in the \cc{}.
The performance model is part of the open-source benchmark suite.

\subsection{Offload Model and Heatmap}
To predict if applications can be executed in the \cn{}, we need to verify if the available network and compute resources satisfy an application's data and processing requirements.
Equations~\ref{eqn:local} and \ref{eqn:offload} describe our performance models for local execution on endpoints and offloaded processing on edge or cloud.
Only when the deployment meets all data and processing requirements will the models deem viable execution in the \cn{}, marked as 1 in the equations.

We assume long-running data processing applications are used with predictable and periodical offloading patterns, such as the machine learning application used in the benchmark evaluation, so we ignore startup and clean-up overheads when offloading tasks.
This scenario is common in edge data processing as sensors and other data-generating devices like cameras are in constant use.

\begin{equation}
    \label{eqn:local}
    \mathit{Local} =
    \begin{cases}
        0 & \text{if $(T_{proc} \times R) > (C_{e} \times Q_{e})$} \\
        1 & \text{otherwise}
    \end{cases}
\end{equation}

\begin{equation}
    \label{eqn:offload}
    \mathit{Offload} =
    \begin{cases}
        0 & \text{if $(T_{proc} \times R \times E) > (C_{o} \times Q_{o})$} \\
        0 & \text{if $(T_{pre} \times R) > (C_{e} \times Q_{e})$}           \\
        0 & \text{if $D > B$}                                               \\
        1 & \text{otherwise}
    \end{cases}
\end{equation}
where:\\
\begin{tabular}{ll}
    $T_{proc}$ & Processing time per data element (sec)             \\
    $T_{pre}$  & Preprocessing time per data element (sec)          \\
    $C$        & CPU cores of endpoint $e$ or offloading target $o$ \\
    $Q$        & CPU quota of endpoint $e$ or offloading target $o$ \\
    $R$        & Data element generation rate (Hz)                  \\
    $E$        & Endpoints connected to offloading target           \\
    $D$        & Data generated per second per endpoint (Mbps)      \\
    $B$        & Bandwidth to offloading target (Mbps)
\end{tabular}

The formula captures compute capacity, represented as various calculations involving the CPU capacity $C$ and quota $Q$ at the offload targets, and compute demand, captured as the time to process each data element ($T_{proc}$ and $T_{pre}$) with data element generation rate $P$.
In the simplest terms, whenever demand (workload property) exceeds capacity (infrastructure property), that configuration setup is not viable (marked as $0$, and the red zone in Figure~\ref{fig:heatmap}).
More specifically, with endpoints generating $R$ images per second, each image should be processed, either locally or on an offload target, before the next image is generated.
This guarantees real-time processing as there are no workload queues building up.

Equation~\ref{eqn:offload} also identifies a non-viable case where data generation rates from endpoints are higher than the available bandwidth.
For viable configurations, there is at least one viable processing location: Either locally on the endpoint (Eq.~\ref{eqn:local}), or remotely on a cloud or edge by offloading (Eq.~\ref{eqn:offload}).

\subsection{Verifying Empirical Results}
We focus on two results marked as \textit{A} (endpoint-only) and \textit{B} (offloaded) in Figure~\ref{fig:bench-endpoint}.
We use the parameters from these experiments as reported in Table~\ref{tab:parameters}, including an $R$ of 5.
The equation for point A is:

\begin{equation}
    \label{eqn:local_fill}
    \begin{split}
        \mathit{Local}: & \hspace{0.1cm } 0.11 \times 5 = 1 \times 0.5 \\
        & \hspace{0.72cm } 0.56 > 0.5
    \end{split}
\end{equation}

Therefore, showing that local processing on an endpoint is not viable in this configuration because the endpoint can not keep up with the data generation speed.
The estimated system load is $(0.56 / 0.5) * 100 = 112\%$, a slight difference from the actual system load of 131\% noted as A in Figure~\ref{fig:bench-endpoint}, but still reaching the same conclusion.
For the offloaded scenario with an aggregation of two endpoints:

\begin{figure}[t!]
    \centering
    \includegraphics[width=.75\linewidth]{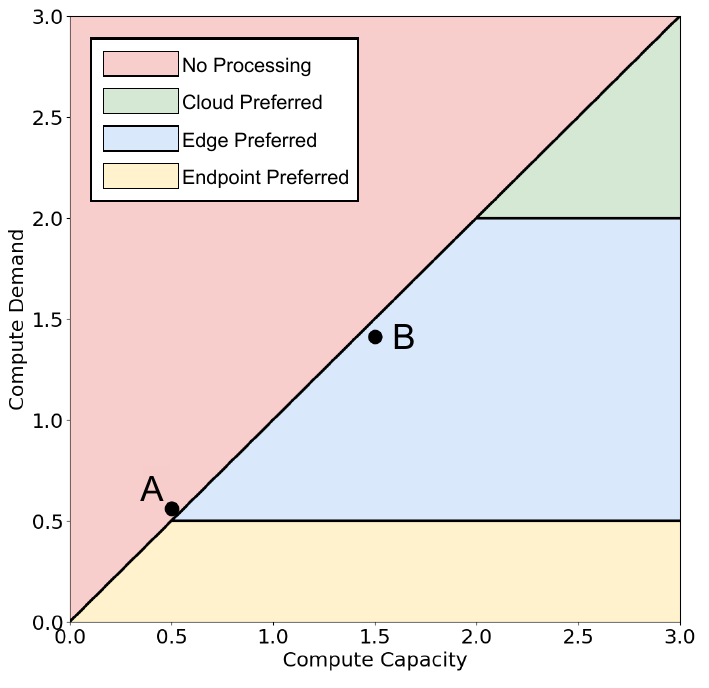}
    \caption{Exploring preferred deployment models for the model examples from Equations~\ref{eqn:local_fill} (A) and \ref{eqn:offload_fill} (B). The division between cloud, edge, and endpoint preferred is configurable and based on computing capacity differences.}
    \label{fig:heatmap}
    \vspace{-0.5cm}
\end{figure}

\begin{equation}
    \label{eqn:offload_fill}
    \begin{split}
        \mathit{Offload}: & \hspace{0.1cm} 0.14 \times 5 \times 2 = 2 \times 0.75 \hspace{0.1cm}\text{and}\hspace{0.1cm}  0.001 \times 5 = 1 \times 0.5 \\
        & \hspace{1.5cm} 1.4 < 1.5                              \hspace{0.88cm}\text{and}\hspace{0.7cm} 0.005 < 0.5
    \end{split}
\end{equation}

The left part of the equation shows that processing can be successfully offloaded to the edge, with an estimated system load of $(1.4 / 1.5) * 100 = 93\%$.
This system load is again not much different from the actual system load of 83\%, denoted as B in Figure~\ref{fig:bench-endpoint}.
All three conditions for offloading are passed, showing the viability of this configuration: Offloading processing to the edge, preprocessing on the endpoint (the right part of Equation~\ref{eqn:offload_fill}), and available bandwidth (2.7 Mbps required compared to 8 Mbps offered).
Hence, our analytical model correctly predicts points A and B, with their position in Figure~\ref{fig:heatmap} indicating deployment viability or non-viability.



\begin{table}[t!]
    \centering
    \caption{Selection of Reference Architectures Mapped to Computing Models for Task Offloading. MC: Mist Computing; EC: Edge Computing; MEC: Multi-access Edge Computing; FC: Fog Computing; MCC: Mobile Cloud Computing. Symbols: \CIRCLE : Present; \Circle : Not present.}
    \label{tab:relatedwork}
    \adjustbox{width=\linewidth}{
        \begin{tabular}{l|ccccc}
    \specialrule{.1em}{.0em}{.0em} 
    Authors                                           & MC      & MEC     & EC      & FC      & MCC   \\
    \specialrule{.1em}{.0em}{.0em} 
    Yogi et al.\/~\cite{yogi2017mist}                 & \CIRCLE & \Circle & \Circle & \Circle & \Circle \\
    ETSI~\cite{etsi2022multi}                         & \Circle & \CIRCLE & \Circle & \Circle & \Circle \\
    ECC~\cite{ecc2017edge}                            & \Circle & \Circle & \CIRCLE & \Circle & \Circle \\
    IIC~\cite{iiot2018tseng}                          & \Circle & \Circle & \CIRCLE & \Circle & \Circle \\
    Intel, SAP~\cite{intel2018iot}                    & \Circle & \Circle & \CIRCLE & \Circle & \Circle \\
    OpenNebula~\cite{opennebula2021edge}              & \Circle & \Circle & \CIRCLE & \Circle & \Circle \\
    Sittón-Candanedo et al.\/~\cite{sitton2019review} & \Circle & \Circle & \CIRCLE & \Circle & \Circle \\
    Qinglin et al.\/~\cite{qi2019smart}               & \Circle & \Circle & \CIRCLE & \CIRCLE & \Circle \\
    Willner et al.\/~\cite{willner2020toward}         & \Circle & \Circle & \CIRCLE & \CIRCLE & \Circle \\
    Mahmud et al.\/~\cite{mahmud2018cloud}            & \Circle & \Circle & \Circle & \CIRCLE & \Circle \\
    OpenFog Consortium~\cite{openfog2017openfog}      & \Circle & \Circle & \Circle & \CIRCLE & \Circle \\
    Pop et al.\/~\cite{pop2021fora}                   & \Circle & \Circle & \Circle & \CIRCLE & \Circle \\
    Dinh et al.\/~\cite{dinh2013survey}               & \Circle & \Circle & \Circle & \Circle & \CIRCLE \\
    \specialrule{.05em}{.0em}{.0em} 
    \textbf{\Cc{} (this work)}               & \CIRCLE & \CIRCLE & \CIRCLE & \CIRCLE & \CIRCLE \\
    \specialrule{.1em}{.0em}{.0em} 
\end{tabular}

    }
\end{table}

\section{Related Work} \label{sec:relatedwork}
We have discussed past efforts in building isolated computing models in \Sec{sec:refarch}.
To the best of our knowledge, only previous work from Qinglin et al.~\cite{qi2019smart} and Willner et al.~\cite{willner2020toward} has discussed the possibility of combining different computing models, however, this is limited to only edge and fog computing.
Table~\ref{tab:relatedwork} summarizes previous efforts (and us, the last row) that cover different models in their analysis.

We design and implement a deployment and benchmarking framework for the \cc{} to explore the \cn's deployment space.
On infrastructure provisioning and emulation, closest to our work, Symeonides et al.~\cite{symeonides2020fogify} present Fogify, a framework for emulating fog resources, and Hasenburg et al. present a similar framework with MockFog~\cite{DBLP:journals/corr/abs-2009-10579}.
These systems are part of a larger class of cloud, edge, and endpoint resource emulation and simulation frameworks~\cite{taheri2020edge}.
All support limited computing models and therefore are restricted in resources and deployments that can be emulated, unlike our framework, which enables emulation of all resources spawning the entire \cn{}.
Additionally, by emulating virtual machines instead of isolation methods such as containers, our framework allows the configuration and benchmarking of architecture components other than applications.
For example, users can switch between the Kubernetes and KubeEdge resource managers or add a new resource manager; the same applies to operating services.
Table~\ref{tab:relatedwork2} shows the limitations of related work compared to our framework.

On benchmarking continuum resources and systems, closest to our work, Kimovski et al.~\cite{kimovski2021cloud} propose a benchmarking framework spawning edge, cloud, and fog; we add to it DeFog~\cite{mcchesney2019defog} and DeathStarBench~\cite{gan2019death}.
These benchmark tools present a larger class of benchmarking tools for cloud, edge, and endpoint resources~\cite{varghese2021survey}.
We improve upon these systems by offering more control over resource allocation (e.g., physical machines, virtual machines, containers) and deployment models (e.g., mist computing, edge computing, etc.) to benchmark and by allowing greater customization of compute and network resources.
We argue that the coupling of infrastructure emulation and benchmarking that our framework offers is key in efficiently exploring the design space of \cn{} deployments and is not present in any related benchmark frameworks (the first two columns in Table~\ref{tab:relatedwork2}).

\begin{table}[t]
    \centering
    \caption{Comparison of Characteristics of Selected Emulation and Benchmarking Frameworks for Cloud and Edge.}
    \label{tab:relatedwork2}
    \adjustbox{width=\linewidth}{
        \begin{tabular}{l|cc|ccc|c}
    \specialrule{.1em}{.0em}{.0em} 
                                    & \multicolumn{2}{c|}{Infrastructure}                           & \multicolumn{3}{c|}{Benchmark}                                              & This        \\
    Characteristic                  & \cite{symeonides2020fogify}   & \cite{DBLP:journals/corr/abs-2009-10579}   & \cite{kimovski2021cloud}  & \cite{mcchesney2019defog} & \cite{gan2019death} & work        \\
    \specialrule{.1em}{.0em}{.0em} 
    Considers cloud resources       & \CIRCLE                       & \CIRCLE                       & \CIRCLE                   & \CIRCLE                   & \CIRCLE             & \CIRCLE     \\
    Considers edge resources        & \CIRCLE                       & \CIRCLE                       & \CIRCLE                   & \CIRCLE                   & \CIRCLE             & \CIRCLE     \\
    Considers endpoint resources    & \CIRCLE                       & \CIRCLE                       & \CIRCLE                   & \CIRCLE                   & \CIRCLE             & \CIRCLE     \\
    Considers network resources     & \CIRCLE                       & \CIRCLE                       & \CIRCLE                   & \CIRCLE                   & \CIRCLE             & \CIRCLE     \\
    Configurable compute resources  & \CIRCLE                       & \CIRCLE                       & \Circle                   & \Circle                   & \Circle             & \CIRCLE     \\
    Configurable networks           & \CIRCLE                       & \CIRCLE                       & \Circle                   & \Circle                   & \Circle             & \CIRCLE     \\
    Configurable resource managers  & \Circle                       & \Circle                       & \Circle                   & \Circle                   & \Circle             & \CIRCLE     \\
    Configurable operating services & \Circle                       & \Circle                       & \Circle                   & \Circle                   & \Circle             & \CIRCLE     \\
    Application benchmarking        & \CIRCLE                       & \CIRCLE                       & \CIRCLE                   & \CIRCLE                   & \CIRCLE             & \CIRCLE     \\
    Supports all computing models   & \Circle                       & \Circle                       & \Circle                   & \Circle                   & \Circle             & \CIRCLE     \\
    \specialrule{.1em}{.0em}{.0em} 
\end{tabular}

    }
\end{table}

On modeling the performance of \cc{} systems, closest to our work, Majeed et al.~\cite{majeed2020modelling} do performance modeling for workload offloading in the fog; we add to it work on network modeling from Ali-Eldin et al.~\cite{alieldin2021hidden}.
These works provide detailed performance models for specific \cn{} deployments, unlike our first-order model, which can be applied to many models.
Furthermore, our performance model is accompanied by an \cc{} benchmark that helps to quickly iterate on application deployment configurations.

\section{Conclusion and Ongoing Work} \label{sec:conclusion}

In this paper, we have made a case for unifying various past, present, and emerging computing models into a single unified \cc{}.
To accomplish this, we provide a detailed analysis of 17 computing models, identify their unique characteristics, and unify them under a \ra{}.
The \ra{} provides a conceptual framework for a systematic exploration of the deployment design space at the \cn{}.
We then provide two unique instantiations of the \ra{} with machine learning and industrial IoT \cn{} workloads.
To enhance the exploration guidelines, our \ra{} is accompanied by a deployment framework and first-order analytical model that help \cn{} developers to reason about their workloads with strong theoretical and engineering foundations.
We are working to enhance the framework with cloud infrastructure support, energy modeling, and automatic ML-driven exploration capabilities.
The effort to develop the \ra{} is a part of \org{}, and the framework and model are open-source, and under active development: \url{https://github.com/atlarge-research/continuum}.
\section*{Acknowledgement}
This work is funded by NWO TOP project OffSense (OCENW.KLEIN.209), the EU GraphMassivizer project (No. 101093202), the Swedish Research Council (Vetenskapsr{\aa}det) -- project ``PSI'' (No. 2020-05094), and the Knowledge Foundation (KKS) -- project ``SACSys'' (No. 20190021).

\bibliographystyle{IEEEtran}
\bibliography{refs.bib}

\end{document}